# INCREASING THE SELECTIVITY OF THE EARLY NOTIFICATION BIOSENSOR SYSTEM UNDER INFLUENCE OF ACOUSTIC VIBRATION

A.N. Grekov, N.A. Grekov, K.A. Kuzmin, S.S. Peliushenko

1 Institute of Natural and Technical Systems, RF, Sevastopol, Lenin St., 28

The paper presents the results of a study of the impact of acoustic and vibration signals on Black Sea mussels, and determines the necessary technical characteristics of vibration sensors. A method has been developed based on the analysis of the time interval recorded by a valve motion sensor in the form of a monotonically decreasing function after the response of a mussel colony to the influence of a vibroacoustic signal. The method makes it possible to eliminate the calibration of the opening value of the mussel valves at the stage of manufacturing and setting up the biosensor system, as well as to control and determine false positives or incomplete opening of the valves of individual mussels when exposed to point stimuli. A structural and functional diagram of the developed experimental setup is presented.

**Introduction**. When developing a marine automated biosensor early warning system signaling pollutants and assessing the possible danger of this water to aquatic and other organisms, suitable organisms were identified that met the following criteria: be typical for the area, have a high number, live in a given place for a number of years, have the ability to be used in natural conditions and have a response to chemical or physical influence. Black Sea mussels meet these criteria.

However, mussels are multi-sensory biological systems that actively respond not only to pollutants, but also to other factors. Therefore, the disadvantages of the biological warning system, as well as other well-known "early warning systems", are the response thresholds caused by the influence of various external interfering factors, such as changes in temperature, salinity (electrical conductivity), illumination, waves, currents and noises affecting the behavioral and/or physiological reactions of aquatic organisms. As a result, natural changes in environmental factors can lead to the same changes in parameters (causing stress) as anthropogenic (toxic) impacts, which inevitably provokes the generation of false alarm signals; the system becomes environmentally inadequate.

In practice, it has been established that mussels (for example, P. Bernhardus and M. Edulis) respond to vibration, and their spectral sensitivity lies in the range from 5 to 410 Hz. The full methodology is presented in [1, 2]. This spectral range may include the noise of passing ships, driving piles into the seabed, waves hitting a buoy or coastal structures, and a number of other impacts. It was found that mussels are able to respond even to minor periodic fluctuations in environmental factors: for example, temperature - by 0.1 ° C (in winter), salinity - by 0.2‰. An analysis of works [3–10] devoted to the effects of acoustic vibrations on mussels and their reaction showed that they do not reflect issues related to the infrasound range, and also did not consider the calibration of the opening value of mussel valves at the stage manufacturing and setting up the biosensor system, false positives or incomplete opening of the valves of individual mussels when exposed to point stimuli, including fouling and the influence of protein threads (byssus), have not been studied. The need to develop a methodology that takes into account the listed factors and improve the selectivity of the response of mussels is the goal of this work.

Prompt identification of the effects of external interfering factors influencing the behavioral and/or physiological reactions of aquatic organisms and preventing the generation of false alarm signals is an urgent task in the development of an automated early warning biosensor complex for environmental protection. -gical monitoring of the aquatic environment and ensuring continuous measurement of current fluctuations in the functional activity of biosensor organisms both in artificial and in any natural conditions with control and filtering of the effects of external interfering factors, such as temperature, salinity (electrical conductivity ), illumination, signals on Black Sea mussels.

**Main part**. It is known that in bivalves the hearing structure is represented by the sensory organ of the abdominal cavity, which is a mechanosensory receptor highly sensitive to mechanical vibrations. We will determine the effects of acoustic and vibration signals on Black Sea mussels, which will allow us to determine the technical characteristics of acoustic vibration sensors necessary for installation in an automated marine early warning biosystem.

The studies were carried out in laboratory conditions using a developed experimental setup consisting of an aquarium containing a life support system that circulates water and enriches it with oxygen, a time interval generator that produces a generated electrical signal with a certain duty cycle, a vibroacoustic source signal, two-channel voltage and frequency generator, low-frequency power amplifier, mussel colony block connected to a multi-channel mussel activity meter, having mussel attachment points and fixation sensors with a mussel valve opening converter into a code, block telemetry, which has a measuring frame former controller and data transmission nodes.

The developed experimental setup works as follows. The time interval former creates a gate once every 30 minutes to turn on a source of vibro-acoustic signal lasting 2 minutes. An electrical signal of a certain frequency is generated in the source of the vibroacoustic signal, which is fed through a power amplifier to an electroacoustic transducer. The generated vibroacoustic signal from the electro-acoustic transducer spreads through the wall of the aquarium in waves, currents and noises affecting the behavioral and/or physiological reactions of aquatic organisms, and identifying the degree of their deviations from the norm, in order to increase the accuracy and reliability of the indication of environmentally hazardous changes in the environment, primarily toxic pollution. Therefore, the urgent task is to study and evaluate the impact of acoustic and vibration the water, affects the mussel colony and the primary vibration-acceleration transducer, in which it is digitized.

The structural and functional diagram of the experimental setup is shown in Fig. 1.

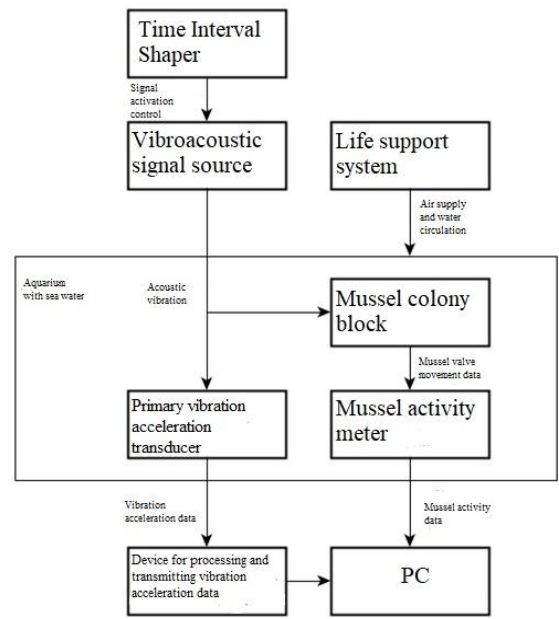

**Fig. 1.** Structural and functional scheme of the experimental setup

Fig. Figure 2 shows a photograph of the experimental setup. The experimental setup consists of an aquarium with sea water, in which there is a colony of mussels, attached to a meter for the motor activity of mussel valves and a hydroacoustic measuring channel. The hydroacoustic measuring channel includes a three-axis accelerometer MPU-9250 mounted on the mussel valve and a digital measuring transducer STM32 Nucleo.

The audio signal source includes a JD6600 generator, a power amplifier and a JBL Stadium 122SSI electroacoustic transducer. The generator is controlled according to the selected mode using a time interval shaper. The life support system circulates water and enriches it with oxygen. The parameters of the movement of mussel valves and real acoustic signals recorded in the aquarium are synchronously sent to the workstation for processing and analysis.

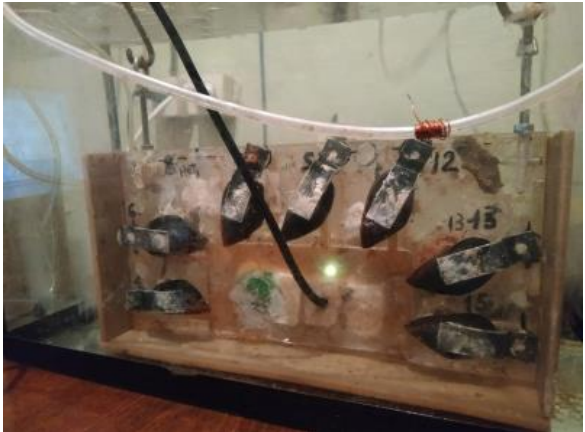

**Fig. 2.** General view of the experimental setup: 1 – plate with a magnet, 2 – mussel, 3 – electronics unit

Basic conditions of the experiments. Sea mollusks (Black Sea mussel) ranging in size from 3 to 5 cm were collected in the area of Sevastopol at a depth of 3-5 m; Black Sea water in a volume of 20 liters was poured into an aquarium where a colony of mussels was located.

General physical parameters of the environment during experiments with mussels:
- air temperature: 24÷27°C;
- water temperature in containers: 22.9÷24.3°C;
- illumination in the air (day/night): 800÷2000/0÷8 lux;
- Water pH: 7.44÷7.95 rel. units;
- date: 08/01/2022–08/21/2022.

Each experiment with different frequencies affecting mussels lasted two days. First day: collecting water and shellfish, installing shellfish in attachment points and setting up hardware. Night and day recordings were carried out continuously for two days.

During the tests, the frequencies given in table were used. 1. A vibroacoustic signal was applied to a colony of mussels with an interval of 30 minutes and a duration of 2 minutes. Additionally, the sound pressure measured by the hydrophone varied from 95 dB and below.

Table 1. Date of the experiment and frequencies of vibroacoustic impact signals

| Date | Frequency Hz |
|---|---|
| 03.08.2022 | 100 |
| 05.08.2022 | 200 |
| 09.08.2022 | 300 |
| 12.08.2022 | 33 |
| 16.08.2022 | 400 |
| 19.08.2022 | 166 |
| 21.08.2022 | 20 |

The results of primary processing of measurement data when exposed to signals of different frequencies and the response of mussel valves are presented in Fig. 3 a, b, c (a – frequency 100 Hz, b – 20 Hz, c – 300 Hz).

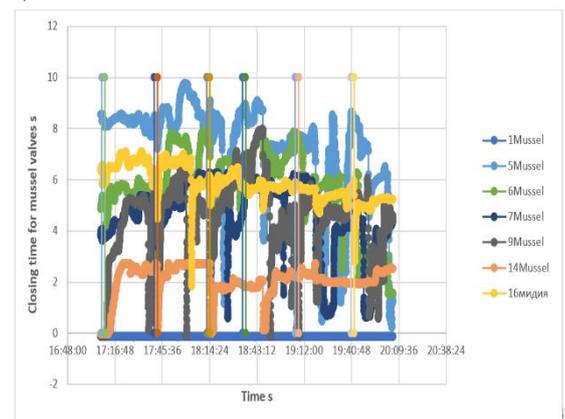

**Fig. 3 a.** The results of the primary processing of measurement data when exposed to a signal with a frequency of 100 Hz and the response of mussel valves

Figure 3a shows the reactions of mussels at a frequency of exposure of 100 Hz. Vertical lines indicate the inclusion of a vibration acoustic signal source with a duration of 2 minutes, which is started every 30 minutes. After exposure to the vibration signal, mussels react differently to it, the amplitude and duration of the closure of the flaps change with a different range. Similarly, a graph is constructed for the reaction of mussels at a frequency of 20 Hz (Fig. 3b)

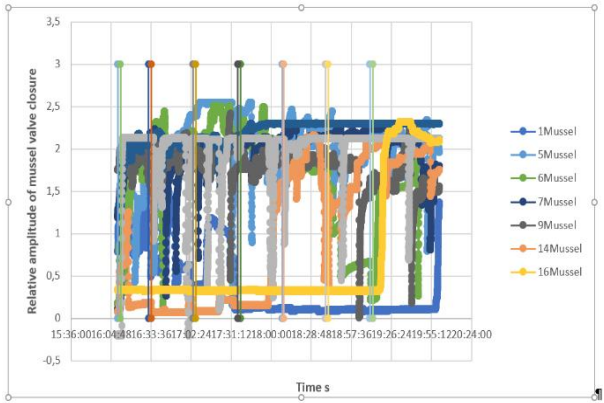

Fig. 3 b. The results of the primary processing of measurement data when exposed to a signal with a frequency of 20 Hz and the response of mussel valves

For clarity, in Fig. Figure 3c shows a fragment with an extended response duration of mussels at a frequency of 300 Hz, where it is clear that the amplitude and time of closure of the mussel valves differ significantly.

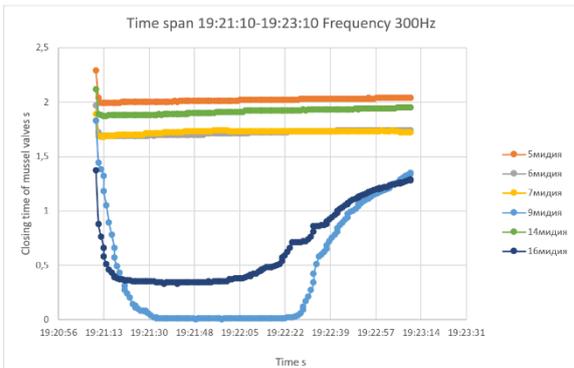

**Fig. 3 в.** A fragment of the graph with an extended duration of the mussel reaction when exposed to a signal with a frequency of 300 Hz

Analysis of the presented signals on the graphs (Fig. 3 a, b, c) in this form is difficult, and sometimes even impossible, since the amplitude of the closure of mussel valves is analyzed, which depends on the initial calibration carried out before the experimental study. The variability of the amplitude of opening of mussel valves depends on the biological characteristics of the individuals themselves, which respond to the food supply, on the oxygen content in the water, on fouling and the influence of protein threads (byssus). To eliminate the above factors, a new approach to the results of processing and analysis was needed. A method was developed that makes it possible to eliminate the calibration of the opening value of the mussel valves at the stage of manufacturing and setting up the biosensor system, as well as to control and determine false positives or incomplete opening of the valves of individual mussels when exposed to point stimuli. The essence of the developed method is that after a colony of mussels is exposed to various stimuli, the time of simultaneous closure of the valves of all mussels, recorded by the valve movement sensor in the form of a monotonically decreasing function, is analyzed, or rather the value of the time interval, which was defined as the difference between the beginning moments the movement of mussel valves and its stop or their complete closure.

As a result of the analysis using the technique of determining time intervals, the dependences of the closure time of mussel valves at various frequencies were constructed, which are presented in Fig. 4. First, for the analysis and construction of graphs (Fig. 4), data were taken without averaging, obtained after the impact of the first pulses of vibration signals on the mussels and their response. As follows from the graphs, the closing time of all mussels does not exceed 10 s. The exception was mussel number 5, although later the closing time was restored to 10 s. Without going into biological aspects, which are not the purpose of the study of this work, one can only assume that the mussel was subjected to fouling or the influence of protein threads (byssus), which broke off after exposure to lower vibration frequencies.

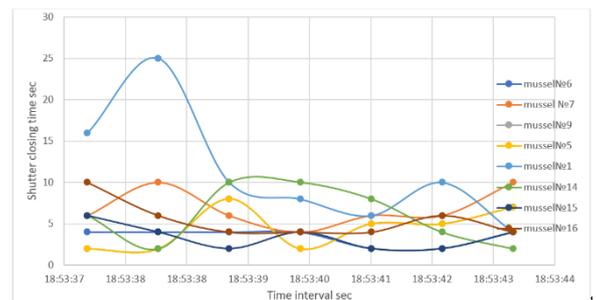

**Fig. 4.** Closing time of the valves of individual mussels depending on the frequency of exposure to the vibration signal

Using the obtained data presented in Fig. 4, let us average the time of transition of the valves of all mussels to the closed state in each frequency range

$$\bar{z} = \sum_{i=1}^{N} z_i,$$

where $\bar{z}$ is the average time of transition of the valves to the closed state; N – number of valve closures.

The results of averaging are presented on the graph (Fig. 5 a), from which it follows that on average for all mussels in the specified frequency range there is a valve closing reaction and the maximum value of the average value of the transition to the closed state does not exceed 2.5 s. To assess the degree of influence of the vibration signal in the frequency range below 20 Hz, we will approximate the series of data presented on the graph (Fig. 5 a) in the range from 20 to 100 Hz by a second-degree polynomial, and extrapolate the values for frequencies below 20 Hz. The found dependence of the time of transition to the closed state on the frequency of the vibration signal has the form

$$z = 0{,}0001\, f^{\,2} + 0{,}0008\, f + 1{,}4801,$$

where z is the time of transition of the valves to the closed state; f – vibration signal frequency.

This operation was performed due to the fact that the electrodynamic converter does not reproduce a signal with a frequency below 20 Hz. A frequency below 20 Hz was excited by the movement of the water surface in the aquarium, the signal was successfully recorded by the primary vibration acceleration transducer.

The resulting reference point from the water disturbance did not go beyond the interpolation curve (Fig. 5 b). This result confirms that mussels respond to infrasonic vibrations in the Hertzian fraction range.

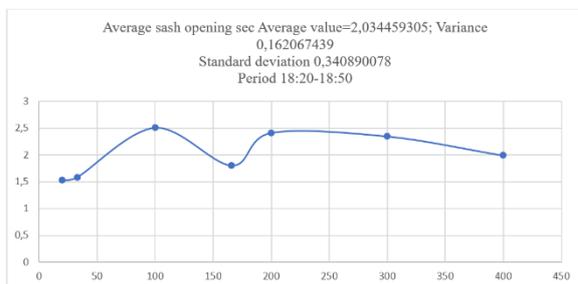

**Fig. 5 a.** The average closing time of all mussels, depending on the frequency of exposure to the vibration signa

$$h = \frac{R}{k}\ ,$$

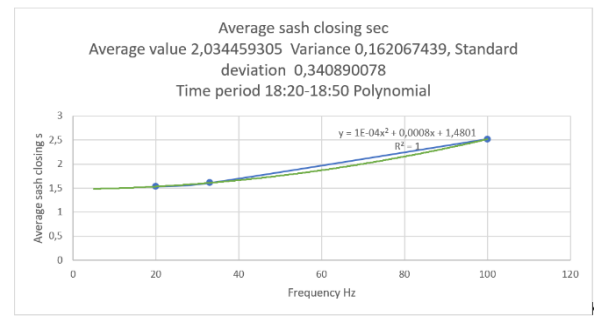

**Fig. 5 б.** Extrapolation curve of the closing time of mussel valves in the infrasonic vibration range

For further analysis of the measurement results, histograms were calculated and constructed for all mussels, but for each frequency separately, and a generalized histogram for all mussels and frequencies. The calculated values for constructing the histogram were obtained after sorting the numerical sample and obtaining the minimum and maximum values for the closure time of the mussel valves. The range of variation was determined and the optimal number of intervals for the closure of mussel valves and the length of the intervals were found

$$R = \chi_{\max} - \chi_{\min},$$

where χmax is the maximum value of the sash closing time; χmin – minimum value of the shutter closing time; R – range of variation.

Optimal number of intervals
$k = 1 + 3{,}22 \cdot \log(n)_{10}$-la Sturgess
where n is the sample length.
Interval length
where k is the number of intervals.
Using the calculated values of all the parameters listed above, we construct a histogram, which is presented in Fig. 6.

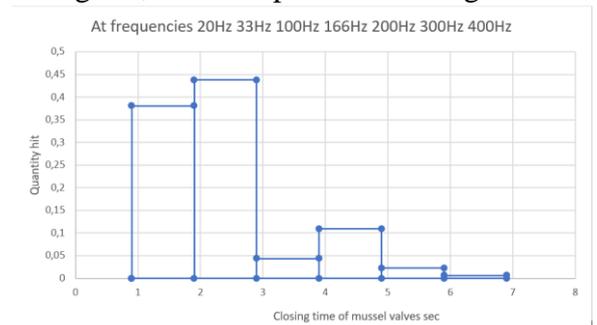

**Fig. 6.** Bar chart

**Conclusion.** As a result of the research, assessments were made of the impact of acoustic and vibration signals of various frequencies, duty cycles and amplitudes on Black Sea mussels, which made it possible to determine the technical characteristics of acoustic vibration sensors required for installation in an automated marine biosensor system. early warning system, and use the developed methodology when processing signals to increase the selectivity of the response to pollutants.

A method has been developed that makes it possible to eliminate the calibration of the opening value of mussel valves at the manufacturing stage and the adjustment of the biosensor system, as well as to monitor and determine false alarms when the valves of individual mussels are not fully opened or closed when exposed to point irritants, including fouling and influence protein threads (byssus).

Bioindication, being simple, allows one to obtain information about biological changes in the environment and draw indirect conclusions about the characteristics of the factor itself. Thus, when assessing the state of the environment, it is desirable to selectively supplement the biological system with channels measuring physicochemical parameters and channels for eliminating false alarm signals, which will allow obtaining qualitative and quantitative characteristics of the marine environment.